\documentclass[prl,aps,12pt,superscriptaddress]{revtex4}
\usepackage{graphicx}
\usepackage{latexsym}
\usepackage{amsmath}
\usepackage{epstopdf}
\begin{document}
\preprint{arXiv:0706.2187}

\title{Algebraic charge liquids}

\author{Ribhu K. Kaul}
\affiliation{Department of Physics, Harvard University, Cambridge MA
02138}

\author{Yong Baek Kim}
\affiliation{Department of Physics, University of Toronto,
Toronto, Ontario M5S 1A7, Canada}

\author{Subir Sachdev}
\affiliation{Department of Physics, Harvard University, Cambridge
MA 02138}

\author{T. Senthil}
\affiliation{ Department of Physics, Massachusetts Institute of
Technology, Cambridge, Massachusetts 02139}

\date{June 2007}
 \maketitle

{\bf High temperature superconductivity of the cuprates emerges upon changing the electron density of an insulator in which the electron spins are antiferromagnetically ordered.
A key characteristic of the superconductor \cite{shen}
is that electrons can be extracted from them at zero energy only if their momenta take one of four specific
values at the `nodal points'. A central enigma has been the evolution of the zero energy electrons in the metallic state between the antiferromagnet and the superconductor, and recent experiments yield apparently contradictory results. Resistance oscillations in a magnetic field \cite{taillosc,cooper} indicate the zero energy electrons lie on elliptical `Fermi pockets', while photoemission \cite{mohit1,mohit2} indicates they lie on arc-like regions.  We describe new
states of matter, called `algebraic charge liquids', which arise naturally
between the antiferromagnet and the superconductor, and reconcile these observations. Our theory also explains the
density of superconducting electrons, and makes
a number of unique predictions for future experiments.}

Soon after the discovery of high temperature superconductivity, Anderson \cite{pwa} presented influential ideas on its connection to a novel type of insulator, in which the electron falls apart into emergent fractional particles which separately carry its spin and charge. These ideas have been extensively developed \cite{pararmps}, and can explain the nodal zero-energy electron states in the superconductor. However, it is now known that the actual cuprate insulators are not of this type, and instead have conventional antiferromagnetic order, with the electron spins aligned in a checkerboard pattern on the square lattice. A separate set of ideas \cite{stripermp} take the presence of antiferromagnetic order in the insulator seriously, but require a specific effort to induce zero energy nodal electrons in the superconductor.

Our theory of algebraic charge liquids (ACLs) uses the the recently developed theoretical framework  of `deconfined quantum criticality' \cite{deccp} to describe the quantum fluctuations of the electrons. We show how this framework naturally combines the virtues of the earlier approaches: we begin with the antiferromagnetic insulator, but obtain electron fractionalization upon changing the electron density. A number of experimental observations in the so-called
`underdoped' region between the antiferromagnet and the superconductor also fall neatly into place.

A key characteristic of an ACL is the presence of an emergent fractional particle which carries charge $e$, no spin, and has Fermi statistics. We shall refer to this fermion as a `holon'. The holon comes in two species, carrying charges $\pm 1$ in its interaction with an emergent gauge field $a_\mu$, where $\mu$ is a spacetime index; this is a U(1) gauge field, similar to
ordinary electromagnetism. However, the analog of the electromagnetic fine
structure constant is of order unity for $a_\mu$, and so its quantum fluctuations have much stronger effects.  We also introduce operators $f^\dagger_{\pm}$ which create holons with charges $\pm 1$.
>From the $f_\pm$ and $a_\mu$
we can construct a variety of observables whose correlations decay with a power-law as function of distance or time in an ACL. These include valence-bond-solid and charge-density wave orders similar to those observed in recent scanning-tunnelling microscopy experiments \cite{kohsaka}.

While the $f_\pm$ carry the charge of the electron in the ACL,
the spin of the electron resides on another fractional particle, the `spinon', with field operator $z_\alpha$ where $\alpha = \uparrow, \downarrow$ is the spin index.
The spinon is electrically neutral, and also carries $a_\mu$ gauge charge.
In an ACL, the spinon is simply related to the antiferromagnetic order parameter;
if $\hat{n}$ is the unit vector specifying the local orientation of the checkerboard spin ordering, then $\hat{n} = z^\dagger \vec{\sigma} z$, where $\vec{\sigma}$ are the Pauli matrices.

The nomenclature `ACL' signals a formal connection to the previously studied
{\em insulating\/} `algebraic spin liquids' \cite{deccp,rantwen,stableu1,motherasl,millisioffe} which have  power-law spin correlations, and of which the deconfined critical point is a particular example. However, the observable properties
of the ACLs are completely different, with the `algebraic' (or, equivalently `critical')
correlations residing in the charge sector.

We begin our more detailed presentation of the ACLs by first considering the simpler
state in which long-range antiferromagnetic order is preserved, but a density $x$ of electrons (per Cu atom) have been removed from the insulator. This is the AF Metal of Fig.~1.
This state has been extensively discussed and well understood in the literature, and its properties have been recently summarized in Ref.~\cite{kaul1} using the same language we use here. Each missing electron creates a charge $e$, spin $S=1/2$
fermionic `hole' (to be distinguished from the spinless `holon') in the antiferromagnetic state. It is known that the momenta of the holes reside in elliptical Fermi pockets centered at the points  ${K}_i = (\pm \pi/2a,
\pm \pi/2a)$ in the square lattice Brillouin zone (of lattice
spacing $a$) - see Fig~2. This is a conventional metallic state, not an ACL, with hole-like zero energy excitations (or equivalently, the hole `Fermi surface') along the dashed lines in Fig~2. In such a state, both the oscillations of the resistance as a function of applied magnetic field (SdH for Shubnikov-de Haas) {\em and\/} emission of electrons by light (ARPES for angle-resolved photoemission) would indicate zero energy electron states at the {\em same\/} momenta: along the dashed lines in Fig~2.

Now let us destroy the antiferromagnetic order by considering the vicinity of the deconfined critical point in the insulator \cite{deccp}. Arguments were presented in Ref.~\cite{kaul1} that we obtain a stable metallic state in which the electron fractionalizes into particles with precisely the quantum numbers of the $f_\pm$
and the $z_\alpha$ described above. This is the holon metal phase: a similar phase
was discussed in early work by Lee \cite{lee89}, but its full structure was clarified recently \cite{kaul1}. Below we will discuss the properties of the holon metal, and of a number of other ACLs that descend from it.

~\\
({\em i\/}) {\bf Holon metal.\/} \\
There is full spin rotation symmetry,
and a positive energy (the spinon energy gap)
is required to create a $z_\alpha$ spinon.
The zero energy holons inherit the zero energy hole states of the AF Metal, and so  they reside along the dashed elliptical pockets in Fig~2,
and these will yield SdH oscillations characteristic of these pockets \cite{ybkim}.
However, the view from ARPES experiments is very different---the distinction between
SdH and ARPES views is a characteristic property of all ACLs. The physical electron
is a composite of $z_\alpha$ and $f_\pm$, and so the ARPES spectrum will have no
zero energy states, and only a broad absorption above the spinon energy gap.
The frequency $F$ of the SdH oscillations is given by the Onsager-Lifshitz relation
\begin{equation}
F = \Phi_0 \mathcal{A} /(2 \pi^2), \label{sdh}
\end{equation}
where $\Phi_0 = hc/(2e)$ is the flux quantum, and $\mathcal{A}$ is the area in momentum space enclosed by the fermionic zero-energy charge carriers. For the holons, this area is specified by the Luttinger relation: there are 4 independent
holon pockets, and the area of each pocket is
\begin{equation}
\mathcal{A}_{\rm holon} = (2 \pi)^2 x/(4 a^2).
\label{ll1}
\end{equation}

~\\
({\em ii\/}) {\bf Holon-hole metal \/}\\
This is our candidate state for the normal state of the cuprates at low hole density (see Fig~1).
It is obtained from the
holon metal when some of the holons and spinons bind to form a
charge $e$, $S=1/2$ particle, which is, of course, the
conventional hole, neutral under the $a_\mu$ charge.
This
binding is caused by the nearest-neighbor electron hopping, and the computation
of the dispersion of the bound state is described in the supplement.
Now the metal has both holons and hole charge carriers, and both
have independent zero energy states, {\em i.e.\/} Fermi surfaces. These Fermi surfaces
are shown in Fig.~2, and {\em both\/} enclosed areas will contribute
a SdH frequency via Eq.~(\ref{sdh}). The values of the areas depend upon specific
parameter values, but the Luttinger relation does yield the single constraint
\begin{equation}
\mathcal{A}_{\rm holon} + 2 \mathcal{A}_{\rm hole} = (2 \pi)^2 x/(4 a^2);
\label{ll2}
\end{equation}
the factor of 2 prefactor of $\mathcal{A}_{\rm hole}$ is due to $S=1/2$
spin of the holes. This relation can offer an explanation of the
recent experiment \cite{taillosc} which observed
SdH oscillations at a frequency of $530 T$.  We propose that
these oscillations are due to the holon states. The holon density may then be inferred to be  $0.076$ per Cu site while the total doping is $x=0.1$. The missing
density resides in the hole pockets, which by Eq.~(\ref{ll2}) will
also exhibit oscillations at the frequency associated with the
area $[(2\pi)^2/(4 a^2)] \times 0.012$. The presence of SdH oscillations
at this lower frequency is a key prediction of our theory which we hope
will be tested experimentally. ARPES experiments detect only the hole Fermi surface,
{\em i.e.\/} the `bananas' in Fig.~2. With a finite momentum width
due to impurity scattering, and the very small area of each banana, the holon-hole
metal also accounts for the arc-like regions observed in current ARPES experiments \cite{shen,mohit1,mohit2}.
Furthermore, because the nearest neighbor electron hopping is suppressed by
local antiferromagnetic correlations, we expect the binding to increase with temperature,
and this possibly accounts for the observed temperature dependence of the arcs \cite{mohit2}.

~\\
({\em iii\/}) {\bf Holon superconductor\/}.\\
Upon lowering the temperature in the holon metal, the holons pair to form a
composite boson which is neutral under the $a_\mu$ charge, and the
condensation of this boson leads to the holon superconductor.
Just as we determined the holon dispersion in the holon metal phase
by referring to previous work in the proximate AF metal, we can determine
the nature of the pair wavefunction by extrapolating from the
state with co-existing antiferromagnetism and superconductivity in Fig~1.
The latter state was studied by Sushkov and collaborators \cite{sushkov1,sushkov2},
and they found that holes paired with $d$-wave symmetry. Hence the AF+dSC
state in Fig~1. We assume that the same pairing amplitude extends into
the ACL obtained by restoring spin rotation symmetry and inducing a spinon
energy gap. The resulting holon superconductor is not smoothly
connected to the conventional BCS $d$-wave superconductor because of the spin gap.
The theory describing the low energy excitations of the holons in the holon
superconductor is developed in the supplement: it is found to be a mathematical structure know as a conformal field theory (CFT). The present CFT has
the U(1) gauge field $a_\mu$ coupled minimally to $N=4$ species of Dirac fermions, $\psi_i$ ($i=1\ldots N$), which are descended from the $f_\pm$ holons after pairing.
Such CFTs have been well-studied
in other contexts \cite{rantwen,stableu1,motherasl,franz}. For us, the important utility of the CFT is that it allows us to compute the temperature ($T$) and $x$ dependence of the density of superfluid electron, $\rho_s$
(measured in units of energy  through its relation to the London penetration depth, $\lambda_L$ by  $\rho_s = \hbar^2 c^2
d/(16 \pi e^2 \lambda_L^2)$, with $d$ the spacing between the layers in the cuprates).
For this, we need the coupling of the
CFT to the vector potential $\vec{A}$ of the electromagnetic field: this is studied
in the supplement and has the form $\vec{j} \cdot \vec{A}$ where $\vec{j}$ is a conserved `flavor' current of the Dirac fermions $\psi_i$. The $T$ dependence of $\rho_s$ is
then related to the $T$ dependence of the susceptibility associated with $\vec{j}$: such susceptiblities were computed in Ref.~\cite{csy}. In a similar manner we found
that as $T \rightarrow 0$ at small $x$
\begin{equation}
\rho_s (x,T) = c_1 x - \mathcal{R} k_B T
\label{rhos}
\end{equation}
where $c_1$ is a non-universal constant and $\mathcal{R}$ is a
universal number characteristic of CFT. Remarkably, such a
$\rho_s$ is seen in experiments \cite{pendepth1,pendepth2,letacon} on the cuprates over a range of $T$ and $x$.
The phenomenological importance of such a $\rho_s (x,T)$
was pointed out by Lee and Wen \cite{leewen97}, although Eq.~(\ref{rhos}) has
not been obtained in any earlier
theory \cite{ivanov}. The holon superconductor provides an elegant
route to this behavior, moreover with $\mathcal{R}$ universal. We
analyzed the CFT in a $1/N$ expansion and obtained
\begin{equation}
\mathcal{R} = 0.4412 + \frac{0.307(2)}{N}
.\label{resr}
\end{equation}

We note that in the cuprates, there is a superconductor-insulator transition
at a non-zero $x=x_c$, and in its immediate vicinity distinct quantum critical behavior of $\rho_s$ is expected, as has been observed
recently\cite{sit}.
However, in characterizing different theories of the underdoped
regime, it is useful to consider the behavior as $x \rightarrow 0$ assuming the superconductivity
survives until $x=0$. In this limit, our present theory is characterized by $d \rho_s /d T \sim
\mbox{constant}$, while theories of Refs.~\cite{ivanov,mohit3} have $d \rho_s /d T \sim x^2$.

~\\
({\em iv\/}) {\bf Holon-hole superconductor.\/}\\
This is our candidate state of the superconductor at low hole
density (see Fig~1).
It is obtained from a pairing instability of the holon-hole
metal, just as the holon superconductor was obtained from the holon metal.
It is a modified version of the holon superconductor, which
has in addition low energy excitations from the paired holes.
The latter will yield the observed V-shaped spectrum in tunneling
measurements. The holon-hole metal will also have a residual metallic thermal conductivity at low $T$ in agreement with experiment. For $\rho_s (x,T)$ we will obtain in addition to the terms
in Eq.~(\ref{rhos}), a contribution from the nodal holes to
 $d \rho_s /dT$. This contribution can be computed using the considerations presented in Refs.~\cite{leewen97,ivanov}: it has only a weak $x$-dependence coming from the ratio of the velocities (which can in principle be extracted from ARPES) in the two spatial
directions. In particular, the holes carry a current $\sim 1$ and not
$\sim x$; the latter is the case in other theories \cite{ivanov,mohit3} of electron fractionalization which consequently have $d \rho_s /dT \sim x^2$.

Perhaps the most dramatic implication of these ideas is the possibility that the superconducting ground state of the underdoped cuprates is not smoothly connected to the conventional superconductors described by the theory of
Bardeen, Cooper, and Schrieffer(BCS). With the reasonable additional assumption that such a BCS ground state is realized in the overdoped cuprates, it follows that our proposal requires at least one quantum phase transition {\it inside} the superconducting dome.

\section{Methods}
We
represent the electron operator\cite{kaul1} on square lattice site $r$ and
spin $\alpha = \uparrow,\downarrow$ as
\begin{eqnarray}
\label{slaves1}
c_{r \alpha} & \sim & f^\dagger_r  z_{r \alpha}~~~~~ r \in A \\
& \sim & \varepsilon_{\alpha \beta} f^\dagger_r z^*_{r\beta}~~~~ r
\in B \label{slaves2}
\end{eqnarray}
where $A,B$ are the 2 sublattices, and $\varepsilon_{\alpha\beta}$ is
the unit antisymmetric tensor. The
$f_r$ are spinless charge-$e$ fermions carrying opposite
$a_\mu$ gauge charge $\pm 1$ on the 2 sublattices. They will be
denoted $f_{\pm}$ respectively.
The effective action of the doped antiferromagnet on the square lattice has the structure
\begin{eqnarray}
\mathcal{S} & = & \mathcal{S}_z + \mathcal{S}_f + \mathcal{S}_{t} \nonumber \\
\mathcal{S}_z & = & \int d\tau \sum_r \frac{1}{g} \left|\partial_\tau z
\right|^2 - \sum_{ \langle rr' \rangle}\frac{1}{g'}\left(z^*_r z_{r'} + \mbox{c.c.}
\right) \nonumber \\
\mathcal{S}_f & = & \int d\tau \sum_{s = \pm} \sum_K
f^{\dagger}_{s} (K) \left(\partial_{\tau} + \epsilon_{K} - \mu_h
\right)
f_{s} (K) \nonumber \\
\mathcal{S}_t & = & \int d\tau \kappa_0 \sum_r c^{\dagger}_r c_r
-\kappa  \sum_{ \langle rr' \rangle} c^{\dagger}_rc_{r'}  +
\mbox{c.c.}, \label{ss}
\end{eqnarray}
where $K$ is a momentum extending over the diamond Brillouin
zone in Fig.~2.
$\mathcal{S}_z$ is the lattice action for spinons.
$\mathcal{S}_f$ describes holons hopping on the same
sublattice, preserving the sublattice index $s = \pm 1$; the holon
dispersion $\epsilon_K$ has minima at the ${K}_i$, and $\mu_h$ is
the holon chemical potential. The first term in $\mathcal{S}_t$
describes an on-site electron chemical potential; the second
to opposite sublattice electron hopping 
between nearest neighbors; the coupling $\kappa$ is
expected to be significantly renormalized down by the
local antiferromagnetic order  from
the bare electron hopping element $t$. We have not included the
$a_\mu$ gauge field in the above action which is easily
inserted by the requirements of gauge invariance and minimal
coupling.   

For small $g$, $g'$, the $z_\alpha$ condense, and we obtain the familiar
AF Metal state with hole pockets (see Fig.~1). For
larger $g$, $g'$ we reach the holon metal phase \cite{kaul1}, in
which the $z$ are gapped.
The $S_t$ term 
leads to two distinct (but not exclusive) instabilities of the
holon metal phase: towards pairing of the $f_{\pm}$ holons and the
formation of bound states between the holons and spinons. These lead, respectively, to the holon superconductor and the holon-hole
metal (and to holon-hole superconductor when both are present).
The $S_t$ term will also significantly modify the spin correlation spectrum, likely inducing incommensurate spin correlations \cite{ss}, but this we will not address here.

Consider, first, the instability due to pairing
between opposite gauge charges, $\langle
f^{\dagger}_+f^{\dagger}_- \rangle \neq 0$ so that the order parameter carries
physical charge $2e$, spin $0$, and gauge charge $0$.
The $a_\mu$ gauge symmetry remains unbroken. The low energy theory of the holon superconductor
has gapless nodal Dirac holons $\psi_i$  coupled to the gauge field $a_\mu$ with the action
\begin{eqnarray}
\mathcal{S}_{\rm holon~superconductor}  &=& \int d\tau d^2 R \Biggl[
 \frac{1}{2e_0^2}\left(\epsilon_{\mu\nu\lambda}\partial_{\nu}a_{\lambda}\right)^2  \nonumber \\
 &~&~~~~~~~~~~~~~~~~~~~~~~~~+ \sum_{i=1}^4
\psi_i^\dagger \left( D_{\tau} -iv_F D_X \tau^x -iv_F D_Y \tau^z \right)\psi_i  \Biggr] \label{ssahl}
\end{eqnarray}
Here $\mu, \nu, \lambda, ...$ are spacetime indices $(\tau,X,Y)$, and
$D_\mu = \partial_\mu - i a_\mu$.
This describes
massless QED$_3$ theory with $N = 4$ species of 2-component Dirac
fermions which flows (within a $1/N$ expansion) at low energies to a stable fixed point \cite{rantwen,stableu1}
describing a CFT.

As the superconductivity arises through
pairing of holons from a Fermi surface of area $\propto x$, the superfluid density  $\rho_s (x,0) \propto x$.
At $T>0$ we need to include thermal excitation of unpaired holons, which are coupled
to the vector potential $\vec{A}$ of the physical electromagnetism by
\begin{displaymath}
\mathcal{H}_A =  \sum_{is}\sum_k \vec A \cdot \frac{\partial \epsilon_{ki}}{\partial \vec k} f^{\dagger}_{iks}f_{iks}
 \equiv \vec j. \vec A
\end{displaymath}
In terms of the Dirac fermions these are readily seen to correspond to 
{\em conserved\/} ``charges'' of Eq.~(\ref{ssahl}). The
susceptibility associated with these charges is $\propto T$ \cite{csy},
and so we obtain Eqn. \ref{rhos}.

Next we consider the $\kappa$-induced
holon-spinon binding in the holon metal which leads to the appearance
of the holon-hole metal at low temperatures.  
$\mathcal{S}_t$ is a momentum dependent attractive contact interaction
proportional
to $\kappa_0 - \kappa \gamma_K $, with $\gamma_K = (\cos(K_x) +
\cos(K_y))$ between a holon and spinon  with center-of-mass
momentum $K$ 
We discuss the energy of a bound holon-spinon composite using a
non-relativistic Schrodinger equation assuming initially that
only a single holon is injected into the paramagnet and ignoring gauge interactions. 
Consider a 
single holon valley (say valley 1). Taking
a parabolic holon dispersion near $\vec K_1$ 
\begin{equation}
E_{f}(\vec k) = \frac{k^2}{2m_h}
\end{equation}
with $\vec k = \vec K - \vec K_1$, and a spinon dispersion
\begin{equation}
E_s (\vec k) = \Delta_s + \frac{k^2}{2\Delta_s}
\end{equation}
centered at $(0,0)$
the energy of the holon-spinon composite will be
\begin{equation}
E_h(\vec k) = \frac{k^2}{2M} + \Delta_s - E_{\rm bind}(\vec k)
\end{equation}
where $M = m_{h}+ \Delta_s$.  Define \begin{equation}
\phi(\vec r) = \left[\begin{array}{l}
\phi_+(\vec r) \\
\phi_-(\vec r)
\end{array}\right].
\end{equation}
with
$\phi_{\pm}(\vec r)$ the wavefunctions of a
composite of a $\pm$-holon and a spinon separated by $\vec r$ and
with center-of-mass momentum $\vec k$.
 This satisfies the Schrodinger equation
\begin{equation}
\left(-\frac{\nabla^2}{2\rho} - (\kappa_0 - \kappa \sigma^x
\tilde{\gamma}(k))\delta^2(\vec r) \right)\phi = -E_{\rm bind}\phi
\end{equation}
where $\rho = m_h \Delta_s/(m_h + \Delta_s)$ and
$\tilde{\gamma}(k) = \gamma(\vec K_1 + \vec k) = -(\sin k_x + \sin
k_y)$. ($\sigma^x$ is a Pauli matrix acting on $\phi$.)
For the momenta of interest, this gives a $\vec k$-dependent hole binding energy
\begin{equation}
E_{\rm bind}^{(1)} \approx V_0 - V_1 \tilde{\gamma}(\vec k)
\end{equation}
with $V_0, V_1 > 0$. Thus the dispersion for $h_1$ becomes
\begin{equation}
E_{h1}(\vec k) = \frac{k^2}{2M} + \Delta_s - V_0 + V_1
\tilde{\gamma}(\vec k)
\end{equation}

The minimum of $E_{h1}$ is at a non-zero energy.  Further
as a function of $k_X = (k_x + k_y)/2$ it is shifted along
the positive $k_X$ direction by an amount $2MV_1 \cos(k_Y)$ with
$k_Y = (-k_x + k_y)/2$.

Considering both holon and hole bands, it is clear that with increasing
doping both bands will be occupied to get a holon-hole
metal. In the full Brillouin zone $h_1$ is at momentum $-\vec K_1$
so that the hole Fermi surface lies entirely inside the diamond
region and has the rough
banana shape shown in Fig. 2. The gauge interaction can now be
included and leads to the properties discussed in the main text.

Correspondence and request for materials to T. Senthil (senthil@mit.edu).
We thank Eric Hudson, Alessandra Lanzara, Patrick Lee, Mohit Randeria, Louis Taillefer, Ziqiang Wang, Zheng-Yu Weng, and Xingjiang Zhou for many useful discussions. This research was supported by the NSF
grants DMR-0537077 (SS and RKK), DMR-0132874 (RKK), DMR-0541988
(RKK), the NSERC (YBK), the CIFAR (YBK), and The Research Corporation(TS).

\newpage

\begin{figure}[t]
\centering \includegraphics[width=6in]{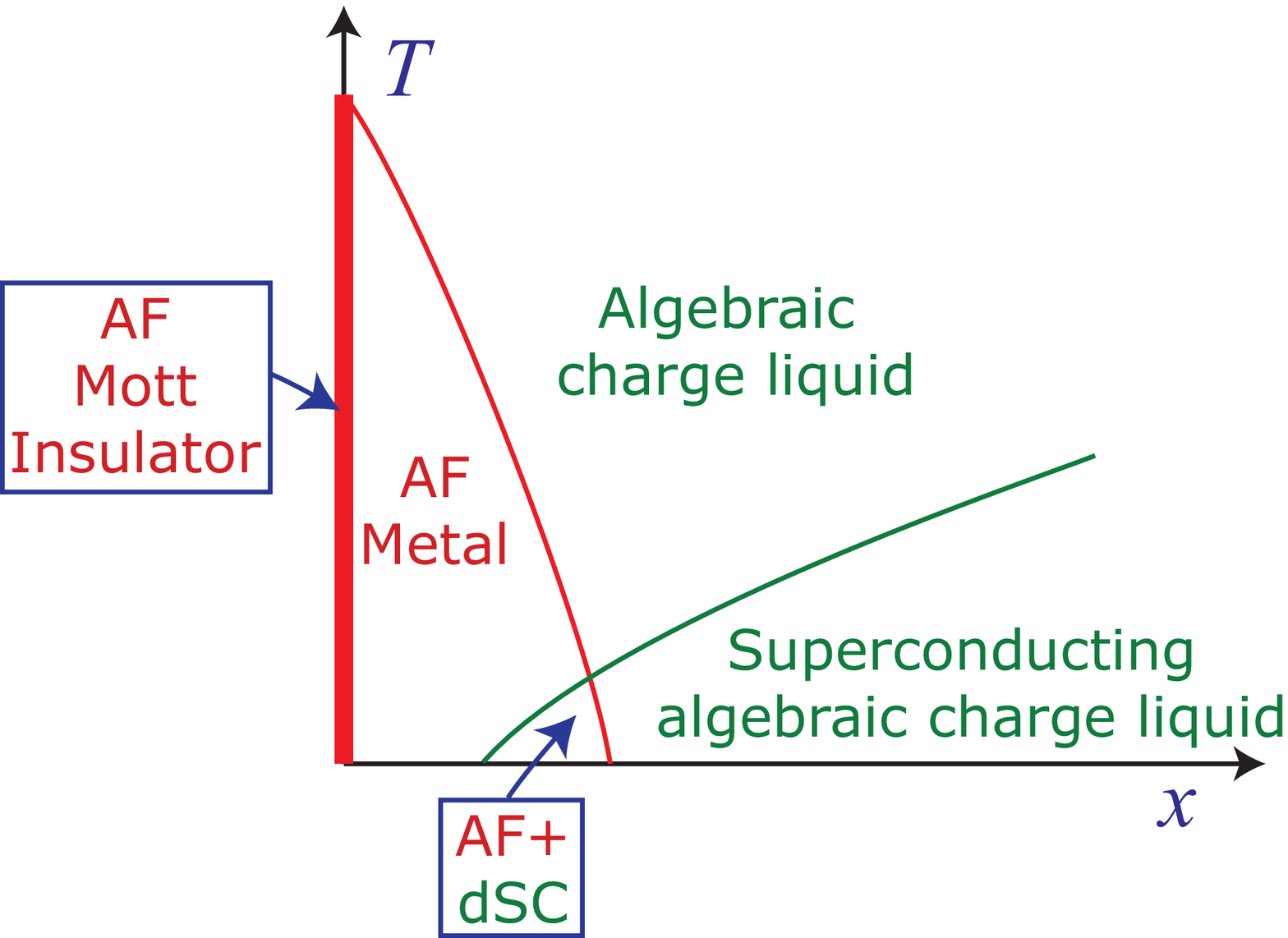}
\caption{{\bf Schematic phase diagram at small $x$}. Here $T$ is the temperature,
and $x$ is the density of electrons (per Cu atom) which have been removed from the insulator.
The phases
labelled ``AF'' have long-range antiferromagnetic order. The AF+dSC state
also has $d$-wave superconductivity and was described in
Ref.~\cite{sushkov1,sushkov2}. For the cuprates, we propose that the
ACL phase above is a holon-hole metal, while the superconducting
ACL is the holon-hole superconductor. The conventional Fermi
liquid metal and BCS superconductor appear at larger $x$, and are
not shown.}
\end{figure}

\begin{figure}[t]
\centering \includegraphics[width=6in]{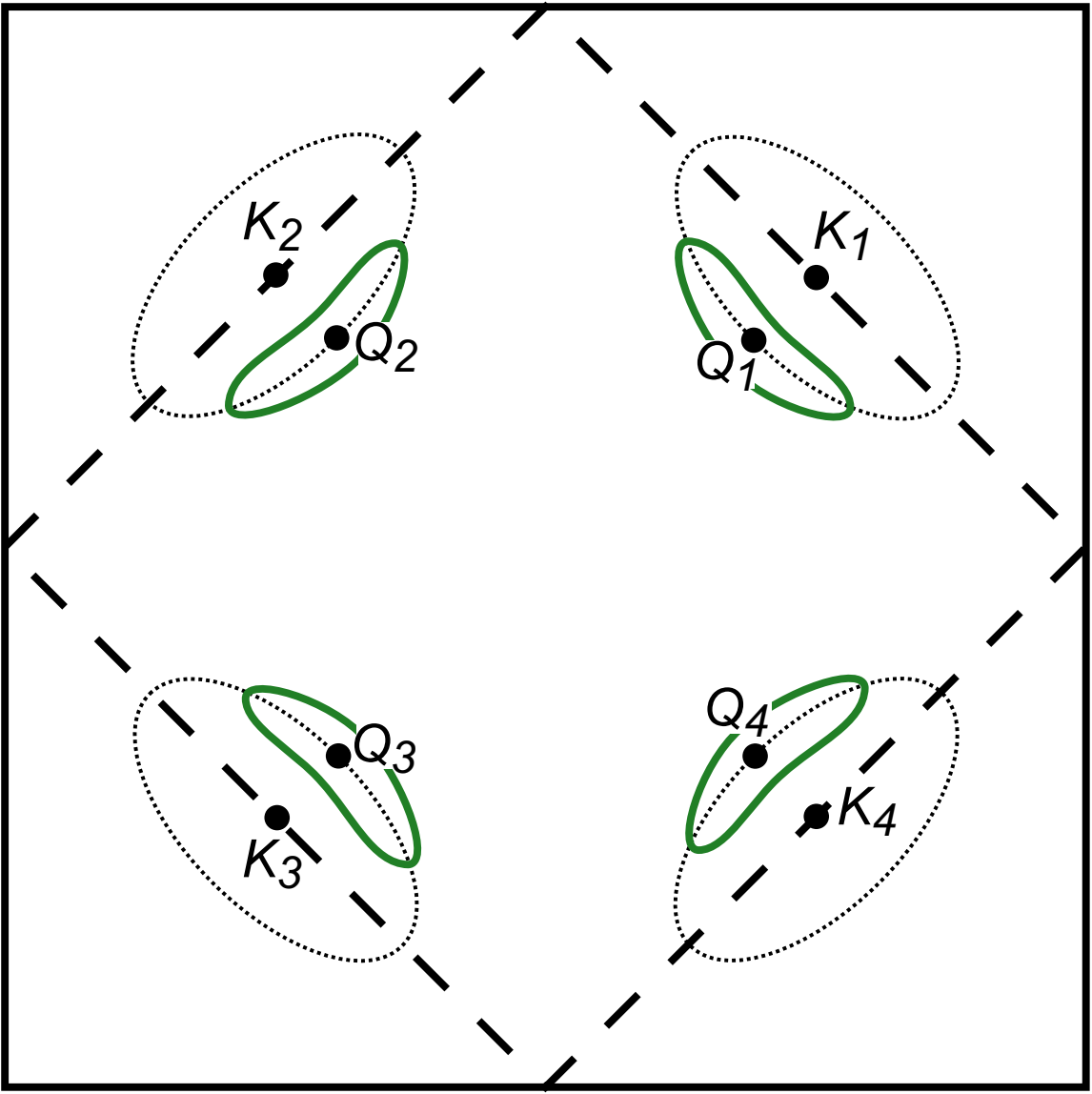}
\caption{{\bf Square lattice Brillouin zone containing the `diamond'
Brillouin zone (dashed line)}. In the AF Metal, only the dashed ellipses
are present, and they represent Fermi surfaces of $S=1/2$, charge $e$ holes which are visible in both ARPES and SdH. In the holon metal,
these dashed ellipses become spinless charge $e$ holon Fermi surfaces, and are
visible only in SdH.
The full line `bananas' are the hole Fermi surfaces present only in the
holon-hole metal, and detectable
in both SdH and ARPES.}
\end{figure}

\end{document}